%% file: main.tex
\documentclass[unnumsec,webpdf,modern,large]{mam-authoring-template}%
\let\textcite\citet
\let\parencite\citep
\let\autocite\citep
\usepackage{graphicx}   
\usepackage{multicol}
\usepackage{amssymb,amsmath}
\usepackage{comment}
\hypersetup{colorlinks=true, citecolor=blue}
\usepackage{hyperref}

\begin{document}

\journaltitle{Microscopy and Microanalysis}
\DOI{DOI HERE}
\copyrightyear{2022}
\pubyear{2019}
\access{Advance Access Publication Date: Day Month Year}
\appnotes{Paper}

\title{Total Variation-Based Image Decomposition and Denoising for Microscopy Images}
\author[1,2,$\ast$]{Marco Corrias}
\author[3]{Giada Franceschi}
\author[3]{Michele Riva}
\author[4]{Alberto Tampieri}
\author[4]{Karin F{\"o}ttinger}
\author[3]{Ulrike Diebold}
\author[5]{Thomas Pock}
\author[1,6,$\ast \ast$]{Cesare Franchini}

\address[1]{\orgdiv{Faculty of Physics and Center for
Computational Materials Science}, \orgname{University of Vienna}}
\address[2]{\orgdiv{Vienna Doctoral School in Physics}, \orgname{University of Vienna}}
\address[3]{\orgdiv{Institute of Applied Physics}, \orgname{TU Wien}}
\address[4]{\orgdiv{Institute of Materials Chemistry}, \orgname{TU Wien}}
\address[5]{\orgdiv{Institute of Computer Graphics and Vision}, \orgname{TU Graz}}
\address[6]{\orgdiv{Department of Physics and Astronomy}, \orgname{University of Bologna}}

\corresp[$\ast$]{Marco Corrias,
\href{mail:1stauthor}{\color{black}{marco.corrias@univie.ac.at}}}
\corresp[\hspace{0.3cm} $\ast \ast$]{Cesare Franchini,
\href{mail:1stauthor}{\color{black}{cesare.franchini@univie.ac.at}}}

\abstract{
Experimentally acquired microscopy images are unavoidably affected by the presence of noise and other unwanted signals, which degrade their quality and might hide relevant features. With the recent increase in image acquisition rate, modern denoising and restoration solutions become necessary.
This study focuses on image decomposition and denoising of microscopy images through a workflow based on total variation (TV), addressing images obtained from various microscopy techniques, including atomic force microscopy (AFM), scanning tunneling microscopy (STM), and scanning electron microscopy (SEM).
Our approach consists in restoring an image by extracting its unwanted signal components and subtracting them from the raw one, or by denoising it.
We evaluate the performance of TV-$L^1$, Huber-ROF, and TGV-$L^1$ in achieving this goal in distinct study cases. Huber-ROF proved to be the most flexible one, while TGV-$L^1$ is the most suitable for denoising.
Our results suggest a wider applicability of this method in microscopy, restricted not only to STM, AFM, and SEM images.
The Python code used for this study is publicly available as part of AiSurf. It is designed to be integrated into experimental workflows for image acquisition or can be used to denoise previously acquired images.
}

\keywords{total variation, image decomposition, denoising, microscopy, AFM, STM, SEM, STEM}
\maketitle

\section{Introduction}
\input{sections/introduction}
\label{sec:intro}

\section{Materials and methods}
\label{sec:mat_met}
\subsection{Experimental images}
\input{sections/experimental_images}
\subsection{Methods}
\input{sections/methods}

\section{Results and Discussion}
\input{sections/results_disc}
\label{sec:results_disc}

\section{Summary and Conclusions}
\label{sec:conclusions}
\input{sections/conclusions}

\section{Data availability statement}
\label{sec:data}
The AiSurf package, including the code used for this study, can be accessed from \url{https://github.com/QuantumMaterialsModelling/AiSurf-Automated-Identification-of-Surface-images}.

\section*{Acknowledgments}
We are grateful to Michael Schmid for insightful discussions and for his critical reading of the manuscript.
We kindly thank Jacob Madsen and Toma Susi for the insights and technical discussions about STEM, which contributed to enriching this work. We thank the technician Karin Withmore for helping us clarify some technicalities of the SEM apparatus used to acquire Fig.~\ref{fig:exp_images}\textcolor{red}{d}. 
This research was funded in whole or in part by the Austrian Science Fund (FWF) 10.55776/F81. For Open Access purposes, the author has applied a CC BY public copyright license to any author-accepted manuscript version arising from this submission.

\section*{Conflict of Interest}
The authors have no conflicts of interest to declare. MC executed the research and made the main contribution to writing the article. GF, MR, AT, and KF contributed to providing and describing the experimental images. UD, TP and CF supervised the research and constantly revised the article. All authors discussed the article during each step, providing their feedback and supervision.



\clearpage
\bibliography{references}


\end{document}


\title{Supplementary material of "Total Variation-Based Image Decomposition and Denoising for Microscopy Images"}

\maketitle


\renewcommand{\thefigure}{S1}
\begin{figure*}[h]
    \centering
    \includegraphics[width=1.\columnwidth]{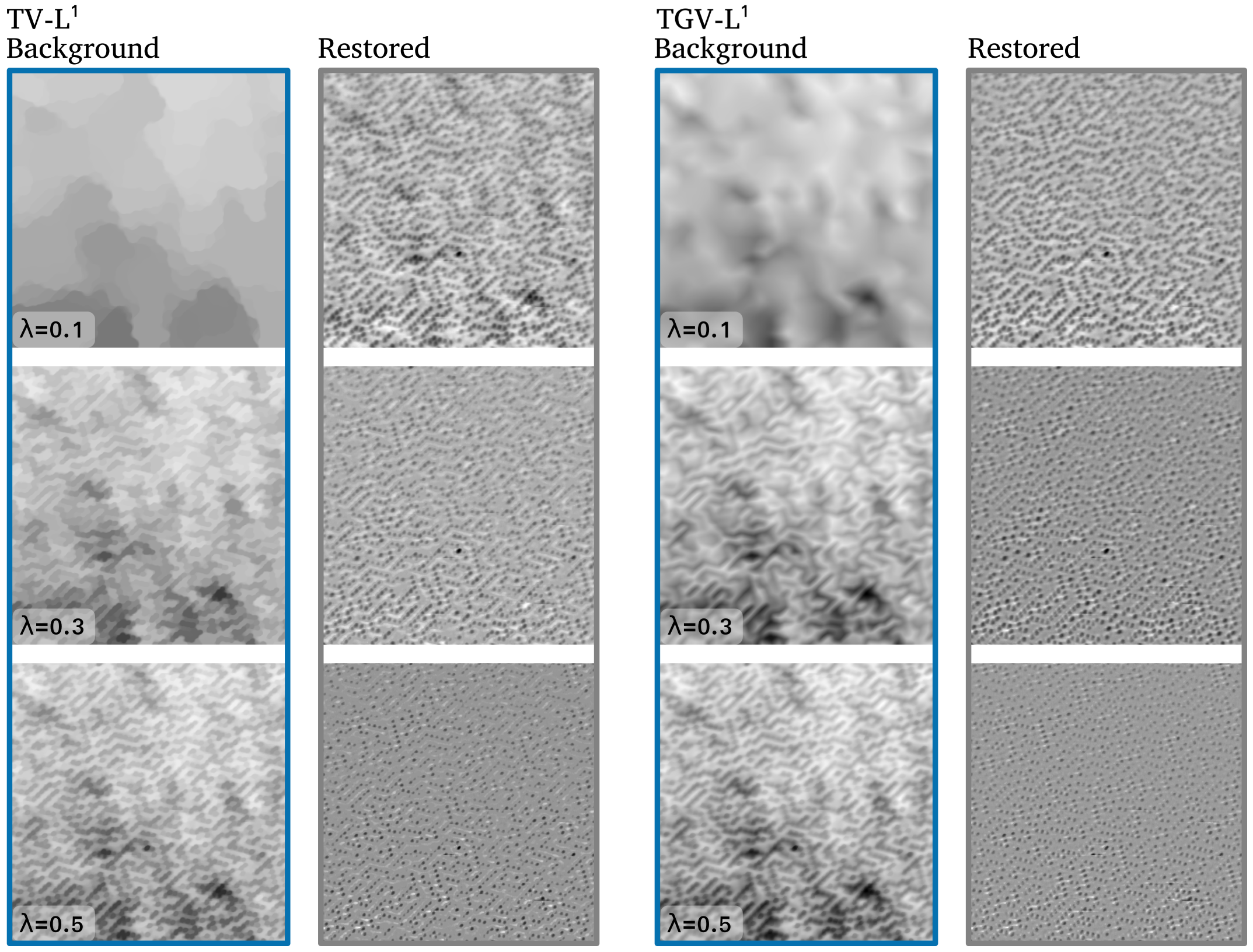}
    \caption{TV-$L^1$ and TGV-$L^1$ applied on the ncAFM of K ions on muscovite mica presented in Fig.~\textcolor{red}{1a}, with different values of $\lambda$.
    Concerning TV-$L^1$, $\lambda=0.1$ is not sufficient to remove unwanted signal, giving a background that poorly represents it. On the other hand, setting $\lambda=0.3$ excessively removes the signal, resulting in a lower-contrast version of the raw image. Further increasing $\lambda$ to 0.5 gives unsatisfying results; thus we conclude that TV-$L^1$ is not suitable for denoising through background subtraction.
    TGV-$L^1$, the second order version of TV-$L^1$, extracts more detailed backgrounds that preserve the original image but still delivers unsatisfying results.
    }
    \label{fig:suppl_mica}
\end{figure*}
\newpage

\renewcommand{\thefigure}{S2}
\begin{figure*}[h]
    \centering
    \includegraphics[width=1.\columnwidth]{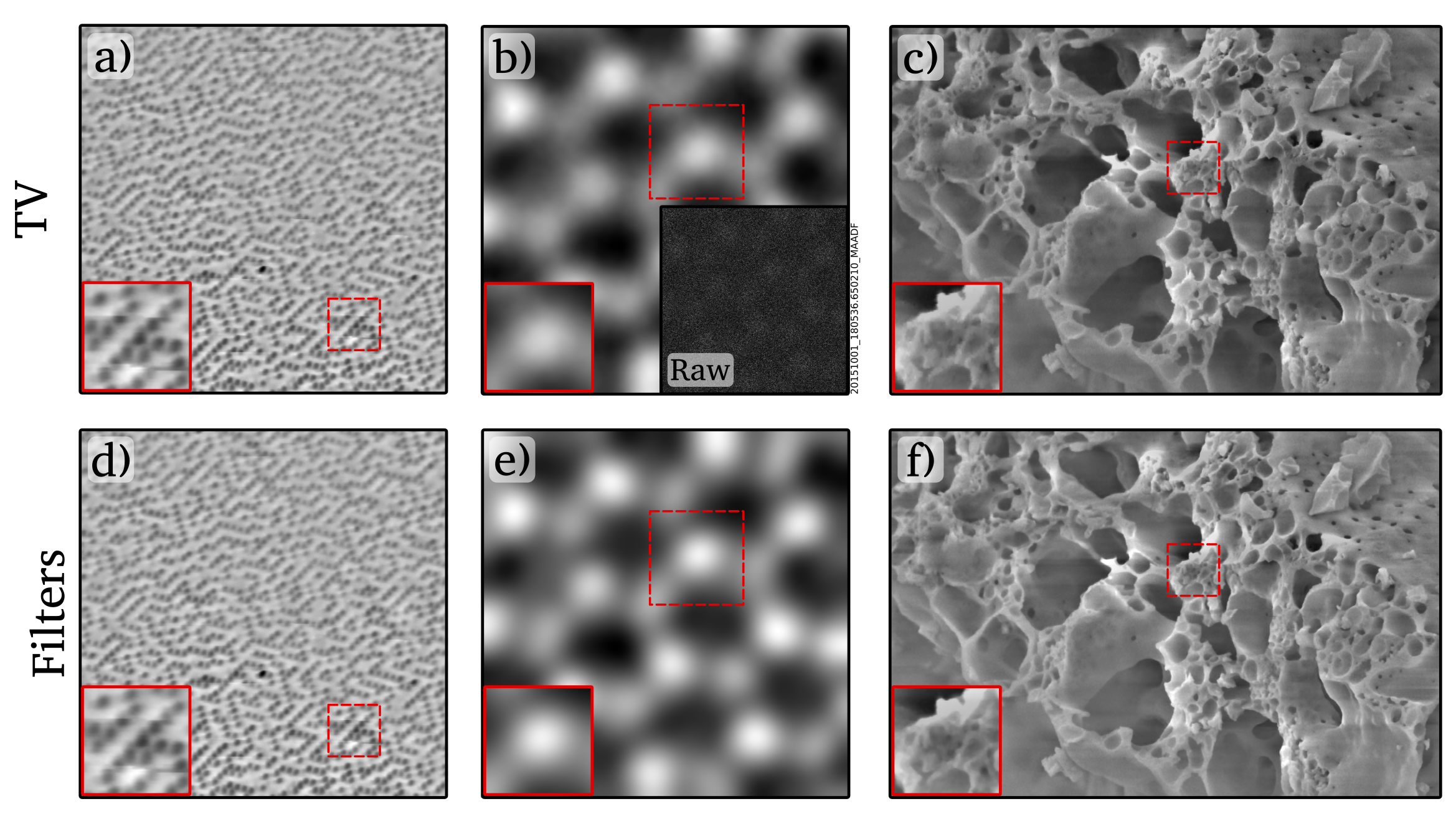}
    \caption{Comparison between images cleaned with total variation ("TV") and filters. (a), (d) restored versions of Fig.~\textcolor{red}{1a}; (b), (e) restored versions of an image taken from the STEM database~\parencite{susi2016database}; (c), (f) restored versions of Fig.~\textcolor{red}{1d}. Panels (a, c) have been restored following the process described in the main text; panel (b) has been smoothed using TGV-L$^1$ with $\lambda=0.1$. Panel (d) has been obtained by applying a Gaussian high-pass filter with $\sigma = 6$ pixels to obtain the background, that then got subtracted to the raw image, and a median filter with $1 \times 3$ kernel. Panel (e) has been obtained by applying a Gaussian low-pass filter.
    Panel (f) has been obtained by applying a high-pass filter and a median filter.
    This figure shows that, apart from minimal differences, the two methods are consistent. The only more noticeable difference is that the image in panel (e) is slightly smoother than the one in panel (b) and shows better depth.
    }
    \label{fig:suppl_comparison}
\end{figure*}

\renewcommand{\thefigure}{S3}
\begin{figure*}[t]
    \centering
    \includegraphics[width=1\columnwidth]{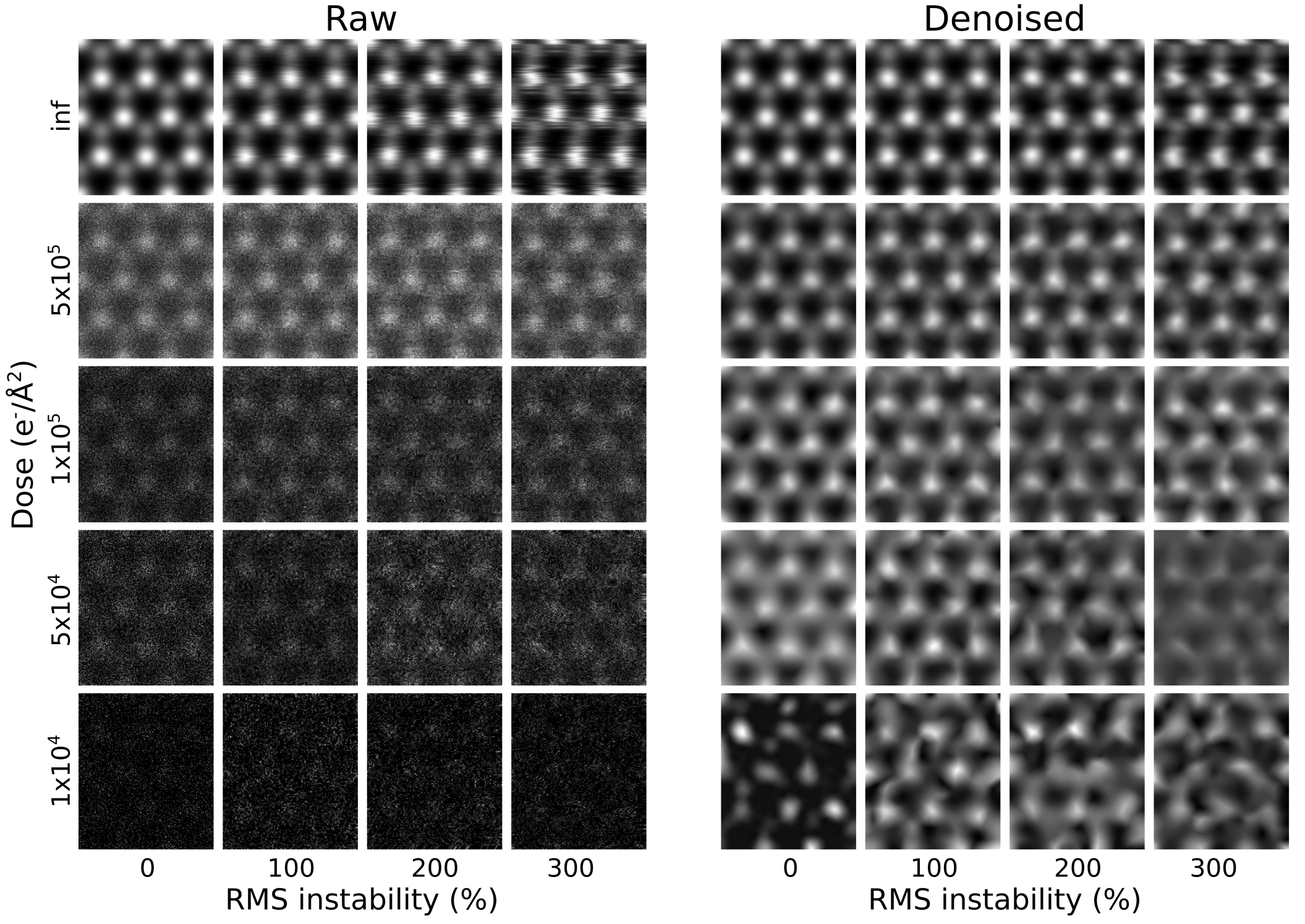}
    \caption{TGV-$L^1$ applied on simulated images of MoS$_2$ using abTEM~\parencite{madsen2021abtem}, with different degrees of scan instabilities and electron doses. The ground truth corresponds to infinite ('inf') electron dose and 0\% scan instability. All the images are denoised with $\lambda=0.1$, RMSE lower than $3 \cdot 10^{-6}$.
    The denoising quality clearly depends on the image quality. For doses higher than $10^5 e^-/$\r{A}$^2$ (i.e. the first three rows), disregarding the scan instability, the denoising quality is good enough to preserve the hexagonal pattern, even though large scan instabilities lead to more distorted patterns. For doses of $5\times10^4 e^-/$\r{A}$^2$ and $10^4 e^-/$\r{A}$^2$ the pattern is almost lost, except for the first two columns of $5\times10^4 e^-/$\r{A}$^2$ dose. Nonetheless, these results assess the good performance of TGV-$L^1$ under challenging noise conditions.
    }
    \label{fig:suppl_stem}
\end{figure*}

\printbibliography 

%% file: sections/introduction.tex
In recent decades, microscopy has experienced significant advancements, particularly with the advent of scanning probe techniques like scanning tunneling microscopy (STM)~\parencite{binnig1982STM, tersoff1985STMtheory} and atomic force microscopy (AFM)~\parencite{Giessibl2003Advances,morita2015noncontact,pavlivcek2017generation}. This innovation has revolutionized our ability to explore surface mechanisms at the atomic scale, marking a pivotal breakthrough in the field.
Notably, the simultaneous performance of STM and AFM has opened new avenues, allowing for a synergistic exploration of both techniques and enhancing the capabilities for comprehensive surface analysis~\parencite{hapala2015simultaneous, durig1986experimental, majzik2013AFMSTM, reticcioli2019interplay}.
Electron microscopy~\parencite{inkson2016EM} is a long-established, versatile technique for small-scale investigations, from micro- to nanoscale.
Techniques like scanning electron microscopy (SEM), transmission electron microscopy (TEM), and scanning transmission electron microscopy (STEM) allow to investigate the topography and chemical composition of materials by analyzing images generated from signals due to electron-sample interactions. \\
Atomically resolved images have the capability to reveal a number of mechanisms and features, encompassing atomic arrangements, molecular adsorption, distribution of defects, and localized charge states such as polarons~\parencite{reticcioli2019interplay, franchini2021polarons}. Experimental investigations at the atomic scale can be further complemented by first-principles calculations, as evidenced by various examples in the literature~\parencite{majzik2013discrimination, spadafora2014identification, stetsovych2015anatase, kraushofer2021singleatom, meier2022COOB, Sokolovic14827}. This integrated approach not only enhances our understanding of intricate phenomena but also showcases the synergistic interplay between experimental observations and theoretical predictions.
It is widely acknowledged that, particularly in small-scale investigations, a significant portion of acquired images exhibits undesirable artifacts and noise, necessitating either their discard or the application of a denoising process~\parencite{fan2019denoisingrev, elad2023denoisingrev}. The nature of noise varies based on the experimental technique and may encompass Poisson, Gaussian, salt\&pepper, or mixed forms. Consequently, integrating a denoising step into the experimental workflow becomes a standard practice to enhance the quality and reliability of the acquired data. Other than noise, images can display undesirable artifacts due to, for example, tip changes in the case of AFM and electric charge build-up in the case of SEM. These typically low-frequency components also need to be filtered in older to obtain an artifacts- and noise-free image. \\
For these reasons, a plethora of denoising methods have been developed by the scientific community, which include filters (both in spatial and transform domain), statistical methods and machine learning~\parencite{thakur2021denoising, goyal2020denoising}.
Some of these methods can be partially of fully automated, but occasionally manual filtering might be a more adequate choice. On this regard, popular choices are WSXM~\parencite{horcas2007wsxm} or ImageJ~\parencite{abramoff2004image, collins2007imagej} plugins; manual filtering allows for fully controlled denoising and processing but also requires experienced users to perform it. On the other hand, automated signal denoising and processing might require limited knowledge on the topic but also gives less control over the outcome. \\
In conjunction with advancements in microscopy, there has been a simultaneous increase in the volume of data generated during experiments. The manual processes of image selection, denoising, and information extraction are progressively becoming more time-consuming, thereby creating bottlenecks not only for surface science but also for diverse fields such as biology, medicine, and surveillance. Addressing this challenge is crucial to streamline workflows, enhance efficiency, and fully leverage the wealth of information embedded in the expanding datasets. To tackle these challenges, contemporary solutions are imperative, and innovative automated techniques are currently being introduced~\parencite{kalinin2021automated, kalinin2022MLstem, madsen2018deeplearning, kalinin2023probe, pregowskascanning}. Popular choices involve the use of machine learning like convolutional neural networks~\parencite{ziletti2018insightful, leitherer2023AISTEM, lin2021temimagenet}, autoencoders~\parencite{kingma2019vae, chen2023cae, biswas2023vae, alvarado2023denoising}, and other algorithmic strategies~\parencite{corrias2023automated, nord2017atomap, belianinov2015identification, zhang2020denoisingSTEM}. \\
Among many possible denoising methods~\parencite{wang2020noise2atom, potapov2019PCAdenoise, kaur2018denoising}, a popular choice is the total variation (TV) minimization~\parencite{chambolle2010introduction, chambolle2011firstorder} due to its ability to preserve fine details and edges while robustly removing different types of noise, as it allows solutions with discontinuities in the functional space, usually associated with object boundaries. Many applications of TV were discovered during the last decades, such as motion estimation~\parencite{werlberger2010motion}, segmentation~\parencite{cai2019linkage}, denoising and deblurring~\parencite{jiaming2019deblurring, Taherkhani2016LinearFA}. TV denoising itself has been applied to different fields, from hyperspectral imaging for environmental monitoring, agriculture, urban planning~\parencite{wang2020hyperspectral}, to spectroscopic~\parencite{liao2015rsiTV}, biological~\parencite{chang2012fluorescenceTV} and medical~\parencite{knoll2011tgvMRI, wang2006medicalTV} imaging. TV denoising was also applied in the field of microscopy~\parencite{kawahara2022stemTV, meiniel2018denoisingrev, chang2012fluorescenceTV}.
Another well-studied application of TV is image decomposition. Image decomposition aims to split an image into its components encoding different types of informations, like structural and textural components~\parencite{aujol2006decomp}, oscillatory, harmonic and structural components~\parencite{huska2021decomp}, or in general a sum of its constituent ones~\parencite{fu2023decomp}. In microscopy imaging, this method can be applied for image restoration, assuming that images are corrupted by noise and other unwanted signal parts.
Moreover, total variation minimization is not affected by image boundary effects caused by popular filtering schemes.
\\
In this work we show how TV denoising and image decomposition can be successfully applied to study cases involving different types of noise and topography, with images obtained from STM, AFM, and SEM experiments. Our study gives practical insights into this methodology and shows its wide degree of applicability on microscopy images, demonstrating it is a valid and non-data-hungry alternative to other modern methodologies. Our results suggest a wider applicability of this method in microscopy, restricted not only to STM, AFM, and SEM images. The Python code used for this study is publicly available on GitHub and is part of the AiSurf~\parencite{corrias2023automated} package. 
This code is designed for single- and multiple-image denoising. Importantly, this methodology holds the potential for seamless integration into automated experimental workflows for image acquisition.

%% file: sections/experimental_images.tex
\begin{figure*}[t] 
    \centering
    \includegraphics[width=1.8\columnwidth]{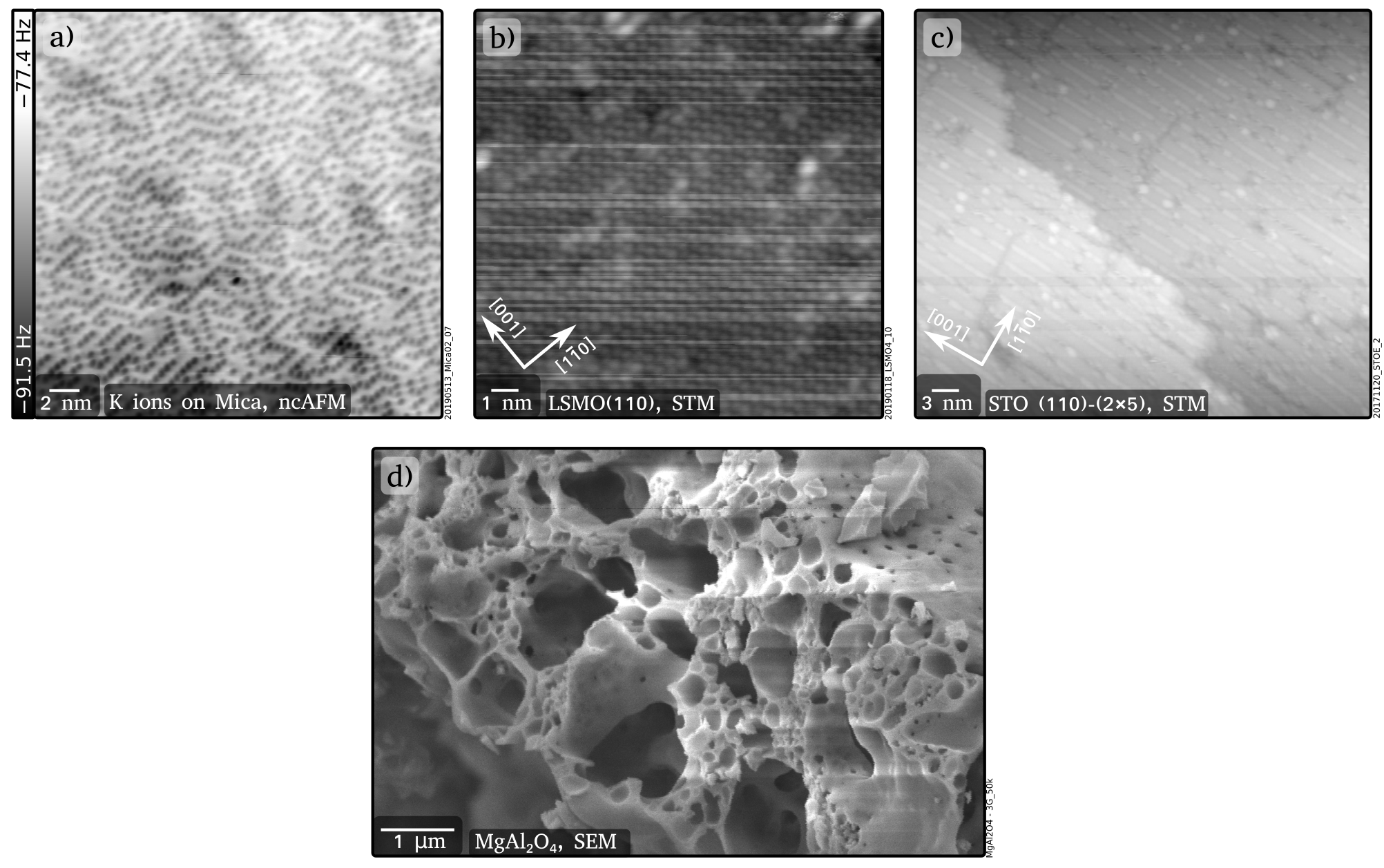}
    \caption{Experimental images discussed in this study. (a) non-contact AFM image of K ions (imaged as dark dots) on muscovite mica. Adapted with permission~\parencite{franceschi2023mica}. \href{https://creativecommons.org/licenses/by/4.0/}{CC BY 4.0}, Nature; (b) STM image of an A-site rich surface of a (110)-oriented lanthanum-strontium manganite (LSMO)~\parencite{franceschi2021LSMO}; (c) STM image of a (2×5)-reconstructed SrTiO$_3$(110) surface~\parencite{riva2019pushing}; (d) SEM image of MgAl$_2$O$_4$.
    }
    \label{fig:exp_images}
\end{figure*}
In this study, four different images will be discussed corresponding to four different study cases, which are collected in Fig.~\ref{fig:exp_images}.
Fig.~\ref{fig:exp_images}\textcolor{red}{a} shows a $27 \times 27$~nm$^2$ non-contact AFM image of the UHV-cleaved surface of muscovite mica, exposing undercoordinated K ions (imaged as dark dots)~\parencite{franceschi2023mica}. Image acquired in constant height from top to bottom with a qPlus sensor~\parencite{giessibl2019qplus} at $4.7$~K with a Cu-terminated tip.
The lower part of the image appears darker (stronger attractive interaction) than the upper one. This is due to thermal drift: while the acquisition was nominally in the constant-height mode, the tip progressively moved closer to the sample surface while scanning from top to bottom. Both the atomic features and the background are affected. Sporadic low-frequency background changes from one scan line to the next are probably due to electrons (charges) jumping between different sites. \\
Fig.~\ref{fig:exp_images}\textcolor{red}{b}, shows a $15 \times 15$~nm$^2$ STM image of an A-site rich surface of a (110)-oriented lanthanum-strontium manganite (LSMO) film obtained by annealing at 700~°C in UHV for 45 minutes~\parencite{franceschi2021LSMO}. $V_t = 2.3$~V, $I_t = 0.05$~nA. The surface features are arranged in zig-zagging rows with a (1$\times$1) periodicity. Sparse, brighter features are assigned to unidentified adsorbates. The sudden, vertical contrast variations (here referred to as 'scratches') are caused by tip changes, likely due to adsorbates attaching and detaching from the tip apex. \\
Fig.~\ref{fig:exp_images}\textcolor{red}{c} shows a $47 \times 42$~nm$^2$ STM image of a (2$\times$5)-reconstructed SrTiO$_3$(110) surface~\parencite{riva2019pushing}. $V_t = 1.9$~V, $I_t = 0.06$~nA. The bright dots aligned along the [001] direction are assigned to Sr atoms, the brighter features appearing in the whole image are unidentified adsorbates. The observed scratches are due to the interaction of the tip with the surface Sr atoms. 
Horizontal scratches are present due to tip changes. They might originate from an adsorbate picked up from the surface, or other phenomena. \\
Fig.~\ref{fig:exp_images}\textcolor{red}{d} shows a 8.2$\times$5.5 $\mu$m$^2$ SEM image of a MgAl$_2$O$_4$ (spinel) particle.
The spinel was synthesized through the combustion method using glycine as a fuel and starting from the corresponding metal nitrates. The obtained solid was ground into powder. The image shows the cleavage surface of the particle, which exposes the material’s inner porosity- a common morphology of solids produced through combustion. The image was taken in high vacuum mode, at 5 kV accelerating voltage and 7.4 mm working distance. Before the analysis, the solid was sputtered with Au:Pd 60:40 to create an 8 nm conductive layer to prevent electric charge build-up on the nonconductive sample, which could alter the secondary electron signal. Despite the sputtering, bright features caused by charging are observed, which could be due to the challenging deposition of the conductive layer on the irregular, porous surface. \\

%% file: sections/methods.tex
\subsubsection{Total variation}
In 1992, Rudin, Osher, and Fatemi introduced the concept of total variation for image denoising~\parencite{rudin1992ROF}, which served as a prototype for many other models that followed~\parencite{chambolle2011firstorder, bredies2010tgv}.
The ROF model is a convex optimization problem with total variation as regularizer and a quadratic data fidelity term. In a discrete setting, for sufficiently smooth functions, the ROF model is defined as:
\begin{equation}
\label{eq:ROF}
\underset{u}{\mathrm{min}} \left\{ \|\nabla u\|_{2,1} + \frac{\lambda}{2} \|u-f\|^2_2  \right\} \hspace{5pt},
\end{equation}
where $u, f \in \mathbb{R}^{m \times n}$ are the sought solution and the input image respectively, 
$ \| \nabla u \|_{2,1} = \sum_{i,j} |(\nabla u)_{i,j} |_2 $, $ |(\nabla u)_{i,j} |_2 = \sqrt{((\nabla u)^x_{i,j})^2 + ((\nabla u)^y_{i,j})^2} $
is the $L^{2,1}$ norm of $\nabla u$, $\lambda >$ 0 is a tuning parameter, and $ \| u-f\|^2_2 = \sum_{i,j} |u_{i,j}-f_{i,j}|^2 $ is the squared $L^2$ norm. The term $\| \nabla u \|_{2,1}$ is named the total variation of $u$.
Choosing the total variation as regularizer is a suitable choice for natural images since it allows the solution $u$ to include sharp discontinuities like edges (e.g. object boundaries), offering a balance between feature preservation and noise removal. \\
Unfortunately, signals treated with the ROF model suffer from the so-called 'staircasing effect', where the output signal shows artificial flat areas. Several solutions have been proposed~\parencite{chambolle2011firstorder}, and in this work we will discuss three of them. \\
The TV-$L^1$ model is obtained by substituting the squared $L^2$ norm with the $L^1$ norm in the data fidelity term of the ROF model: 
\begin{equation}
\label{eq:TVl1}
\underset{u}{\mathrm{min}} \left\{ \|\nabla u\|_{2,1} + \lambda \|u-f\|_1  \right\} \hspace{5pt},
\end{equation}
where $\|u-f\|_1 = \sum_{i,j}|u_{i,j} - f_{i,j}|$.
This simple change brings considerable advantages: this model is contrast invariant, more effective against salt\&pepper noise, and preserves small features better than the ROF model also due to a minimal staircasing effect~\parencite{chan2005tvl1, nikolova2004variational}.
The Huber-ROF model~\parencite{zadorozhnyi2016huber} is another variant of the ROF model. It is obtained by substituting the $L^1$ norm of the total variation in Eq.~\ref{eq:ROF} with the Huber norm:
\begin{equation}
\label{eq:huber}
|\nabla u|_\alpha =
\begin{cases}
    \frac{|\nabla u|^2}{2\alpha} & \mathrm{if} ~|\nabla u| \leq \alpha \\
    |\nabla u| - \frac{\alpha}{2} & \mathrm{if} ~|\nabla u| > \alpha \\
\end{cases}
\end{equation}
The parameter $\alpha >$ 0 is introduced to define a tradeoff between the $L^2$ and $L^1$ norms as regularizers. This model allows for smoother solutions with very limited staircasing effect. \\
The last model we discuss is the (second order) Total Generalized Variation $L^1$ (TGV-$L^1$)~\parencite{bredies2010tgv}. The TGV-$L^1$ model consists in solving the problem
\begin{equation}
\label{eq:TGVl1}
    \underset{u,v}{\mathrm{min}} \left\{
    \alpha_1 \|\nabla u - v\|_{2,1} +
    \alpha_2 \|\frac{1}{2} (\nabla + \nabla^T)v \|_{2,1} + 
    \lambda \|u-f\|_1 \right\}  \hspace{5pt},
\end{equation}
where $v$ is a vector field which minimizer value is located between 0 and $\nabla u$, allowing for second order contributions. Qualitatively, the explaination is the following: near the edges $\nabla^2 u$ is larger than $\nabla u$, so in order to minimize Eq.~\ref{eq:TGVl1} having $v\sim0$ is beneficial, and restores the standard TV regularization term; on the other hand, in smooth regions $\nabla^2 u$ will be low so it is beneficial to have $v\sim\nabla u$ in order to penalize the second derivatives.
$\alpha_1$, $\alpha_2$ are control parameters. The first two terms of the sum are the (second order) total generalized variation of $u$, while the third one is the data fidelity term. As this model takes into account also the second derivatives of $u$ (more details can be found at~\parencite{bredies2010tgv}), images denoised with it look smoother than the ones with the above methods, given the absence of staircaise effects.
TV-$L^1$ and Huber-ROF have been implemented according to~\parencite{chambolle2011firstorder}, while TGV-$L^1$ to~\parencite{knoll2011tgvMRI}.


\subsubsection{Workflow}
\begin{figure*}[t]
    \centering
    \begin{minipage}[c]{0.2\textwidth}
        \caption{Flowchart of the proposed workflow. Starting from the raw image \textit{f}, its background \textit{b} is extracted via TV minimization and \textit{u} is calculated via subtraction. If \textit{b} contains signal associated to prominent features' contrast \textit{v} (e.g. from terraces or different terminations), the latter gets extracted from \textit{b} and added to \textit{u}.
        If \textit{u} is affected by noise, the signal can be smoothed, obtaining $u'$. If both cases are present, one must first perform the contrast retrieval and then proceed with the smoothing.
        The red circles highlight the output from each TV minimization.
        }
        \label{fig:flowchart}
    \end{minipage} \hfill
    \begin{minipage}[c]{0.77\textwidth}
        \includegraphics[width=1.\columnwidth]{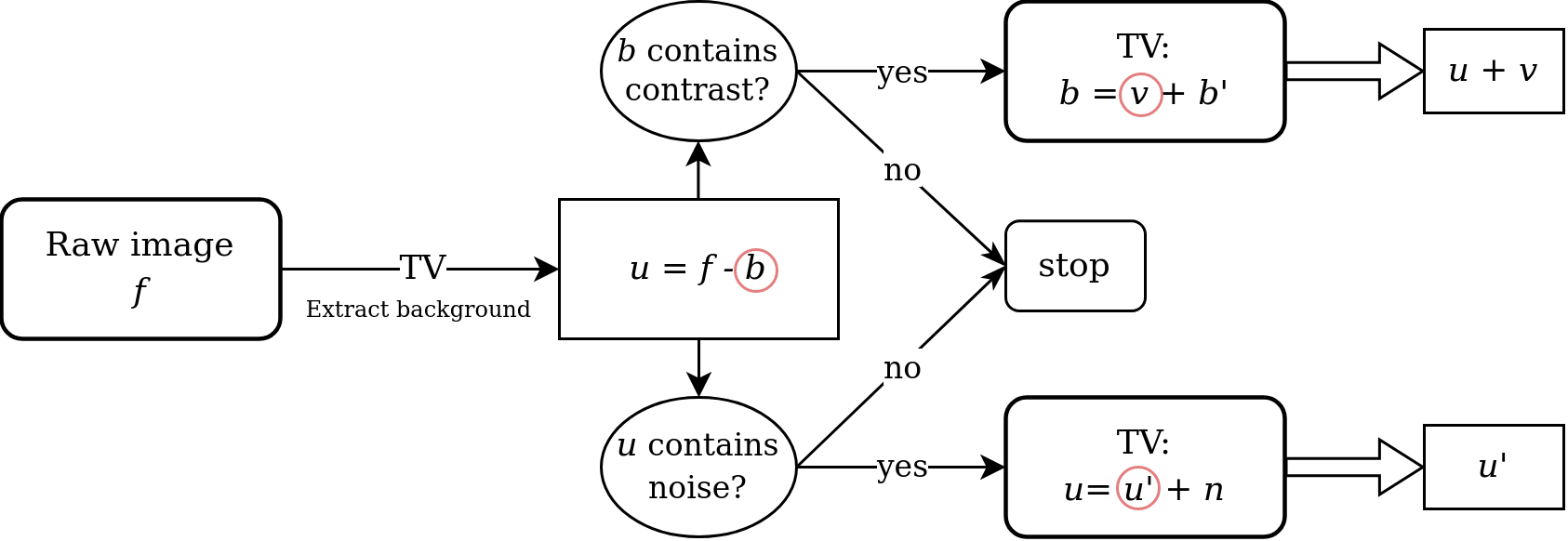}
    \end{minipage}
\end{figure*}

An image \textit{f} can be disassembled as a sum of its components $\{c_1, c_2, ..., c_N\}$~\parencite{fu2023decomp}:
\begin{equation*}
    \label{eq:imgmodel}
    f = \sum_{i=1}^{N} c_i \hspace{1pt}.
\end{equation*}
This definition holds in general, but the problem is ill-posed since the number of components $N$ is not specified a priori. Examples in literature decompose \textit{f} in its structural and textural parts~\parencite{aujol2006decomp}, structural, oscillatory and smooth parts~\parencite{huska2021decomp}, or generally discuss about hierarchical image peeling~\parencite{fu2023decomp}. \\
Given the many possible ways a microscopy image can look like, we refrain from fixing a priori its number of components. Instead, we focus on illustrating through a workflow some cardinal examples and how tackle them. This workflow has been applied to the images discussed in this work, but the user can adapt it to their specific needs.
The main steps of the workflow are summarized in Fig.~\ref{fig:flowchart}.
The first step consists in decomposing the raw image \textit{f} in two parts \textit{b} and \textit{u}: \textit{b} is the background, which includes low-frequency signals like tip scratches and other constrast variations (e.g. from terraces or different terminations); \textit{u} encodes high-frequency components that define the edges of the image, like atomic arrangements, prominent structural features, or even noise. In the simplest of cases, performing this step is sufficient to obtain a clean image \textit{u}. Optionally, a median filter with a $1 \times 3$ kernel is applied to \textit{u} to remove horizontal pixel-thin scratches, frequently present in techniques such as ncAFM and STM. This process has been sufficient to clean the images in Fig.~\ref{fig:exp_images}\textcolor{red}{a, b}. \\
If additional steps are required, we can have two other possible main cases.
If the background \textit{b} also contains relevant signal like the contrast between different terminations or terraces \textit{v}, it is necessary to decompose \textit{b} in two terms to obtain \textit{v}, which will be added to \textit{u} to obtain the final image. 
This strategy has been applied to Fig.~\ref{fig:exp_images}\textcolor{red}{c}. 
If \textit{u} is affected by noise, a second TV minimization can be performed to obtain a smoother, ideally noise-free version of \textit{u}, here named $u'$. Fig.~\ref{fig:exp_images}\textcolor{red}{d} has been cleaned with this procedure. Trivially, if an image is just affected by noise TV minimization can be applied to smoothen it and ideally remove the noise.
Lastly, if an image contains both different-contrast areas and noise, one must first restore the contrast, then smoothen the final result.

%% file: sections/results_disc.tex
\begin{figure}[t]
    \centering
    \includegraphics[width=1\columnwidth]{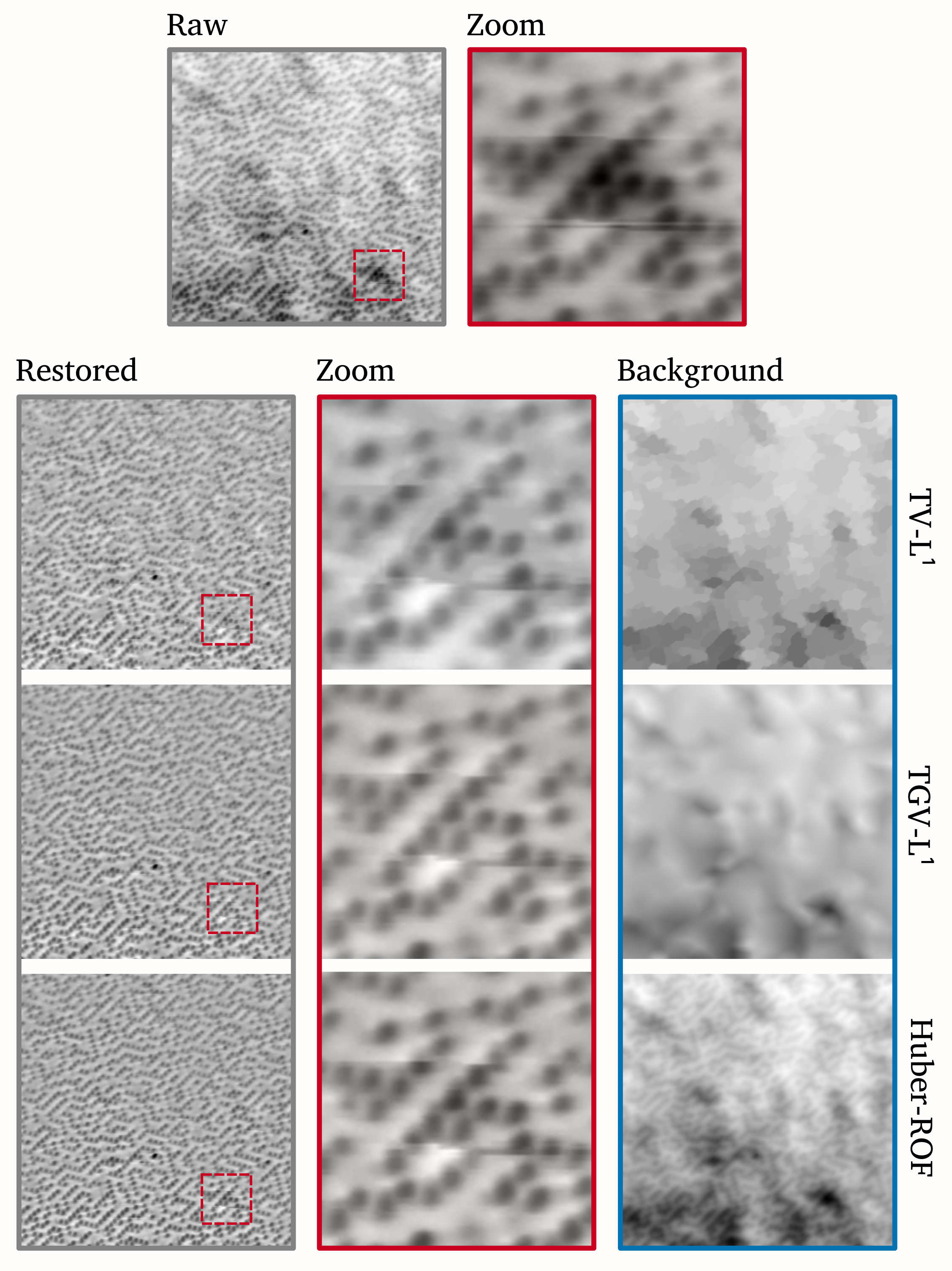}
    \caption{Comparison between the background subtraction carried out with TV-$L^1$, TGV-$L^1$, and Huber-ROF on ncAFM image of K ions on mica presented in Fig.~\ref{fig:exp_images}\textcolor{red}{a}. The Raw image and a zoomed area are compared with the Restored ones obtained from the background subtraction of the raw image; different backgrounds extracted via TV minimization are displayed in the right column. It is evident how differently the three methodologies perform. Huber-ROF is the most flexible method among the three, applicable to all the cases shown in this study, and it performs best in background extraction. The output for TV-$L^1$ has been obtained with $\lambda=0.2$, the one for TGV-$L^1$ with $\lambda=0.1$, and for Huber-ROF with $\lambda=0.5$. A median filter with a $1 \times 3$ kernel has been applied to the restored images for all the three methods.
    \label{fig:comparison}
    }
\end{figure}

This section presents and discusses the outcomes of our study. In the tuning process, each method involved adjusting only the $\lambda$ parameter. The following parameters remained constant throughout the study: $\tau=0.01$ (timestep related to the algorithm), $\alpha_1=1$, and $\alpha_2=2$ for TGV-$L^1$, while $\alpha=0.05$ was consistently used for Huber-ROF. Notably, a more favorable outcome for Fig.~\ref{fig:exp_images}\textcolor{red}{d} a better result was achieved with $\alpha=0.005$. 
All calculations reached convergence with a Root Mean Square Error (RMSE) lower than $5 \cdot 10^{-6}$ between the background of the current iteration and the previous one. \\
In order to comprehend the domain of applicability of the various methods, it is crucial to compare their performances. Figure~\ref{fig:comparison} illustrates the decomposition outcomes for Fig.~\ref{fig:exp_images}\textcolor{red}{a} using three distinct TV methods.
The output for TV-$L^1$ has been obtained with $\lambda=0.2$, the one for TGV-$L^1$ with $\lambda=0.1$, and for Huber-ROF with $\lambda=0.5$.

\begin{figure}[t]
    \centering
    \includegraphics[width=0.65\columnwidth]{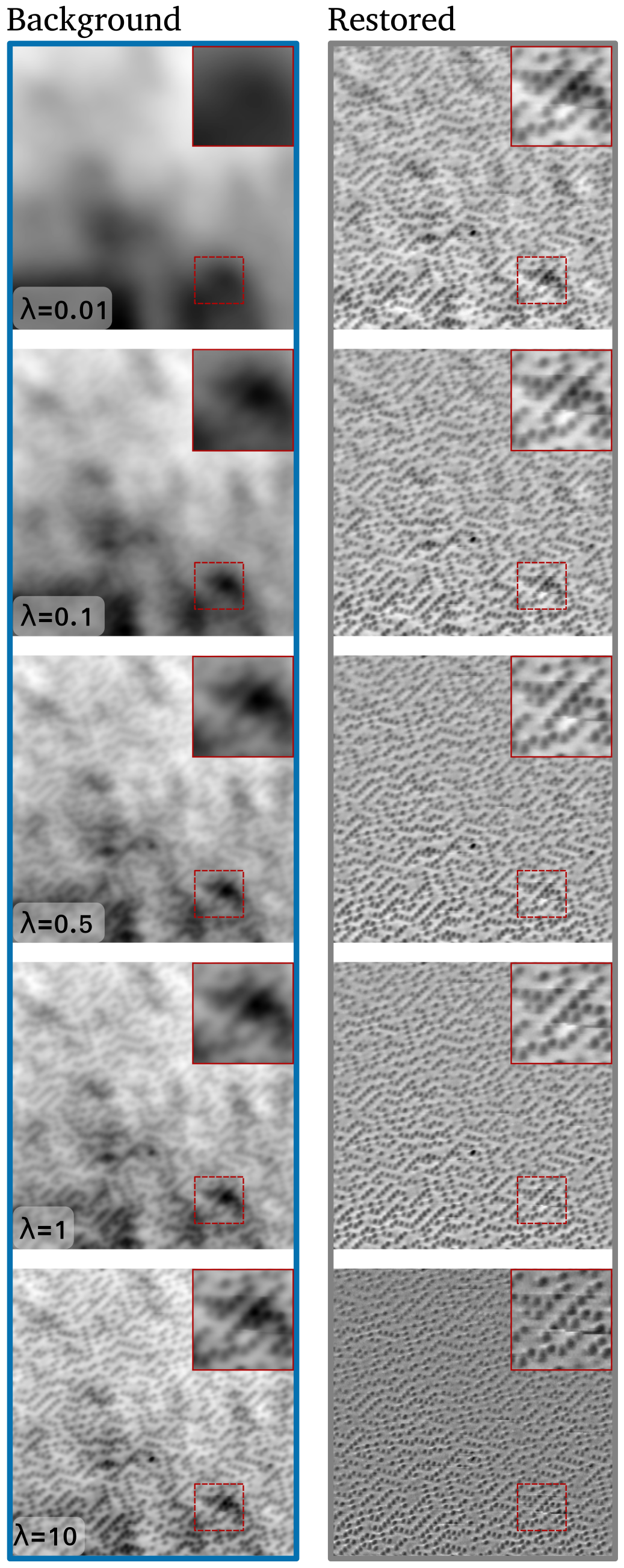}
    \caption{Background and restored image as a function of $\lambda$ using Huber-ROF. Raw image in Fig.~\ref{fig:exp_images}\textcolor{red}{a}. Increasing $\lambda$ allows to filter smaller-sized features. The optimal $\lambda$ value lies between 0.5 and 1, where the background is removed without compromising the contrast. $\lambda=10$ shows that setting this parameter too high excessively reduces the contrast between K ions and the background. This might be beneficial for some tasks that require a clear imaging of the atoms or structures, but for pure restoration purposes it is advised to preserve the original features as much as possible. A median filter with a $1 \times 3$ kernel has been applied to the restored images.
    \label{fig:vslambda}
    }
\end{figure}

Despite being a widely adopted method for TV denoising, the observations from Fig.~\ref{fig:comparison} indicate that TV-$L^1$ is the least suitable among the methods investigated in this study for background subtraction. Even with a carefully chosen $\lambda$ parameter, certain prominent dark spots within the non-uniform background persist, and undesired artifacts appear around the K ions, manifesting as distinct dark dots. As demonstrated in Fig.~\textcolor{red}{S1} in the Supplementary Material, increasing $\lambda$ effectively removes the dark spots but excessively diminishes the contrast between K ions and the bright background, while also introducing new, unwanted artifacts. This behavior may be attributed to the $L^1$ norm as a fidelity term, known for its robustness against outliers, ultimately resulting in backgrounds comprised of spots with constant intensity rather than smoother, higher-detailed solutions. \\
TGV-$L^1$, incorporating second-order terms, facilitates smoother solutions by extracting more detailed backgrounds compared to first-order methods like TV-$L^1$ and Huber-ROF. In this case, no artifacts around the ions are introduced, and the contrast is largely preserved. 
While this outcome is desirable, Fig.~\textcolor{red}{S1} in the \textit{Supplementary Material} indicates that varying $\lambda$ for this method may excessively filter artifacts and noise, potentially yielding unsatisfactory results. For background extraction, achieving a balance in parameter selection is crucial to harness the benefits of TGV-$L^1$ without compromising on the preservation of relevant details. \\
In contrast, as observed in Fig.~\ref{fig:comparison}, Huber-ROF emerges as the most suitable algorithm for the designated tasks among the three. It excels in extracting the background without significantly diminishing the contrast of features, resulting in overall superior outcomes compared to TV-$L^1$ and TGV-$L^1$. Notably, the background achieved with Huber-ROF closely resembles the raw image, encompassing both the non-uniform background and signals from the K ions. While the latter is extracted and the contrast between the bright background and dark features is reduced, the denoised image maintains the sharpness of the features. Adjusting the parameter $\lambda$ also yields more controlled solutions compared to TV-$L^1$ and TGV-$L^1$.
It is also worth mentioning that, depending on the algorithm, parameters, and image used, the computational time varies. TV-L$^1$ and Huber-ROF, being first-order methods, are generally faster than TGV-L$^1$, which is a higher-order one. Increasing lambda decreases the computational time, as the output (i.e. the background) more resembles to the input image. The details and scale of the input image also influence the time factor: larger, more detailed images generally take more time than smaller, less detailed ones. For these reasons, the computational time can go from a fraction of a second to the order of minutes.
\\
Figure~\ref{fig:vslambda} further illustrates the dependence of Huber-ROF on $\lambda$, applied to the same image as in Fig.~\ref{fig:comparison}. 
Increasing $\lambda$ allows to extract features of smaller and smaller scale, going from a larger-scale background to a small-scale signal associated with the atoms; for this reason, it must be selected for background extraction without excessively altering the topographical features of the image (e.g. related to the atomic arrangements). For $\lambda \leq 0.1$ the effects are minimal and do not lead to satisfactory results as the non-uniform background is still present. From $\lambda \geq 0.5$ the dark spots are mostly removed. For $\lambda = 0.5$ we observe how the non-uniform background is removed without affecting the signal coming from the K ions. We argue that the optimal value for $\lambda$ in this case lies between 0.5 and 1 since for the latter the dark halos surrounding the K ions are extracted as well, but their sharpness is preserved. Setting $\lambda = 10$ removes the background but decreases the overall contrast: the single atoms preserve their sharpness, but the excessive subtraction flattens the signal too much, giving the image an unnatural dark-gray background. For purely denoising purposes that aim to preserve the images' original appearance this is negative, but if the task requires highlighting or isolating atomic features like K ions, selecting a high value of $\lambda$ might be beneficial.
Further discussions of parameter selection can be found in~\parencite{chan2005tvl1}. As highlighted in that work, TV-$L^1$ has a clear geometric interpretation related to size selection: this is made evident in Fig.~\textcolor{red}{S1} in the Supplementary material, where the output of different $\lambda$ values is compared. TV-L$^1$ and TGV-L$^1$ qualitatively follow the same behavior, but we have not found an automated way to estimate the parameters $\lambda$ and $\alpha$. Figs.~\ref{fig:vslambda} and \textcolor{red}{S1} however give a good insight on parameter selection, which proved to be sufficient during internal tests with different images.

\begin{figure*}[t]
    \centering
    \includegraphics[width=1.8\columnwidth]{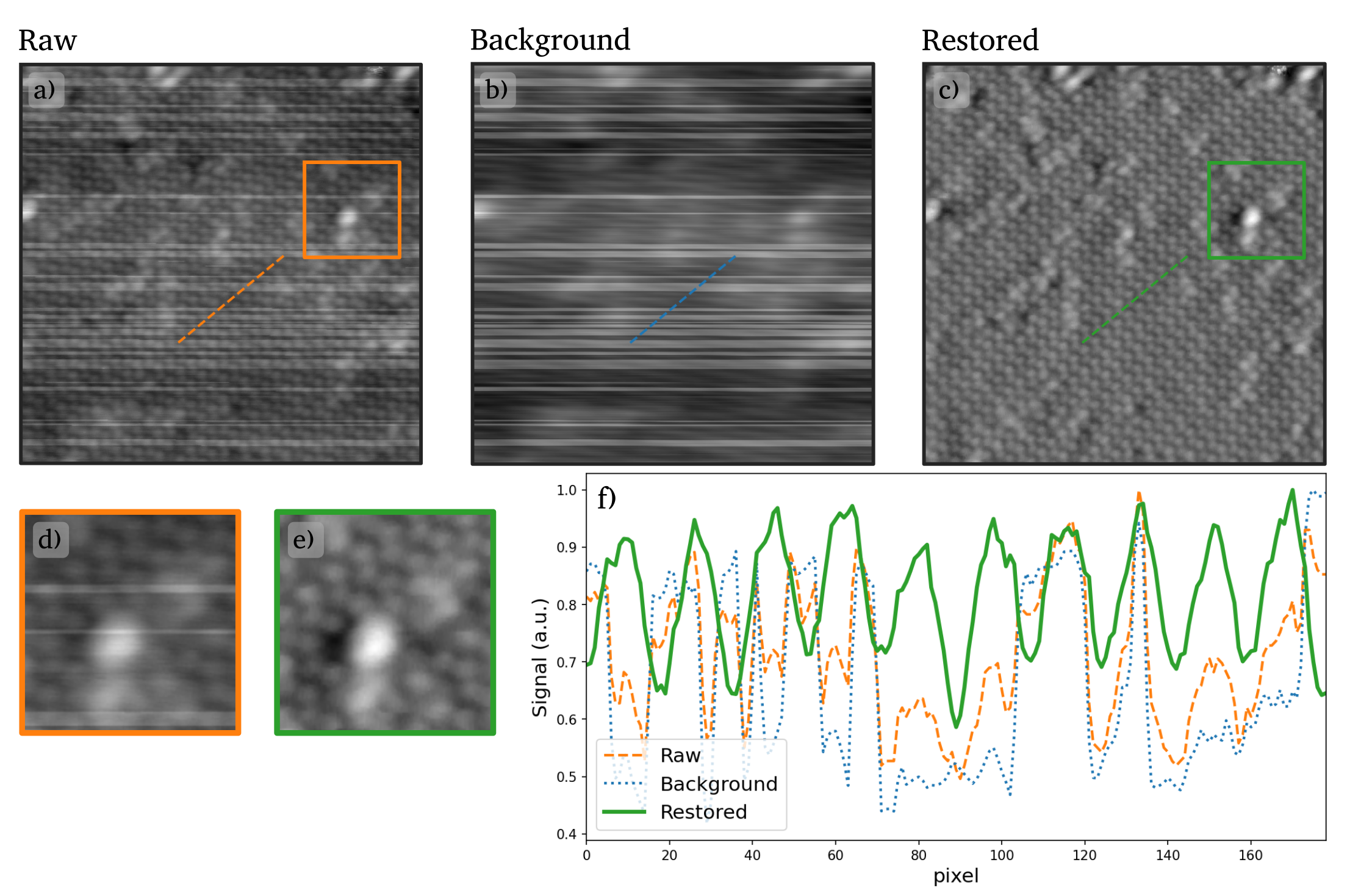}
    \caption{Background subtraction of the STM image of LSMO(110) presented in Fig.~\ref{fig:exp_images}\textcolor{red}{b} via Huber-ROF method, with $\lambda=0.05$. The image is corrupted by visible bright, horizontal lines. (a) raw image overlayed with a dashed line and the region highlighted in panel (d); (b) background extracted via TV minimization overlayed with a dashed line; (c) restored image obtained via background subtraction of the raw image, overlayed with a dashed line and the region highlighted in panel (e); (d), (e) zoomed regions of panels (a), (c) respectively; (f) line profiles relative to the lines overlaying panels (a), (b), and (c), each one normalized to the maximum signal intensity of the image they refer to, for comparison purposes. A median filter with a $1 \times 3$ kernel has been applied to the restored image.
    }
    \label{fig:lineplotLSMO}
\end{figure*}
Figure~\ref{fig:lineplotLSMO} shows the application of Huber-ROF to the STM image of LSMO(110) presented in Fig.~\ref{fig:exp_images}\textcolor{red}{b} and addresses how to operate in case of evident horizontal lines. 
Building on the favorable outcomes observed in our previous results, Huber-ROF is chosen for its capability to extract the background while minimizing the inclusion of signals from atomic features. This ensures the preservation of the original topography of the surface despite the presence of distinct horizontal lines.
In the case of K ions on mica (Fig.~\ref{fig:exp_images}\textcolor{red}{a}), the unwanted signal appears as dark and bright spots non-uniformly distributed, removable by applying TV minimization to both vertical and horizontal directions. In the case shown in Figure~\ref{fig:lineplotLSMO}\textcolor{red}{a} the background is made of perfectly horizontal lines, parallel to the scanning direction; the correct strategy here is to apply TV minimization only along the horizontal direction, ignoring the vertical sharp signal gradients due to the presence of bright lines. Since these lines only slightly vary in intensity along the horizontal direction, they can be seen as 'large-scale objects' and thus require setting a low $\lambda$ to avoid extracting the underneath texture, as discussed in Fig.~\ref{fig:vslambda}. $\lambda=0.05$ has been set for this purpose. The background in panel~b) indeed shows how the lines were accurately extracted, alongside only some signal coming from the bright protrusions. The result in panel~c) is the optimal one from this image, since not only the horizontal lines have been removed, but also the bright atomic features show a higher contrast than before, and their sharpness is preserved; the zoomed areas in panels~d), e) highlight this fact. To visualize how the signal looks in panels~a), b), and c), a line profile is shown in panel~f). A line profile consists of plotting the signal coming from the image as a function of coordinates, highlighted by a line overlaying the image. In our case, this allows to visualize the peak distribution coming from the atomic rows. 
The dashed line associated with the raw image consists of peaks distributed without regularity due to the horizontal lines, covering the periodic pattern underneath. The dotted line, associated with the background, shows that the most prominent peaks are due to the bright lines. The full line shows the periodic signal coming from the atomic rows. 
In summary, the restoration process successfully preserved the topography of the image, retaining the bright atomic rows, even brighter adsorbates, and darker signals originally present. This affirms the effectiveness of the Huber-ROF method, with an observed increase in the contrast of the features.

\begin{figure*}[t]
    \centering
    \includegraphics[width=1.8\columnwidth]{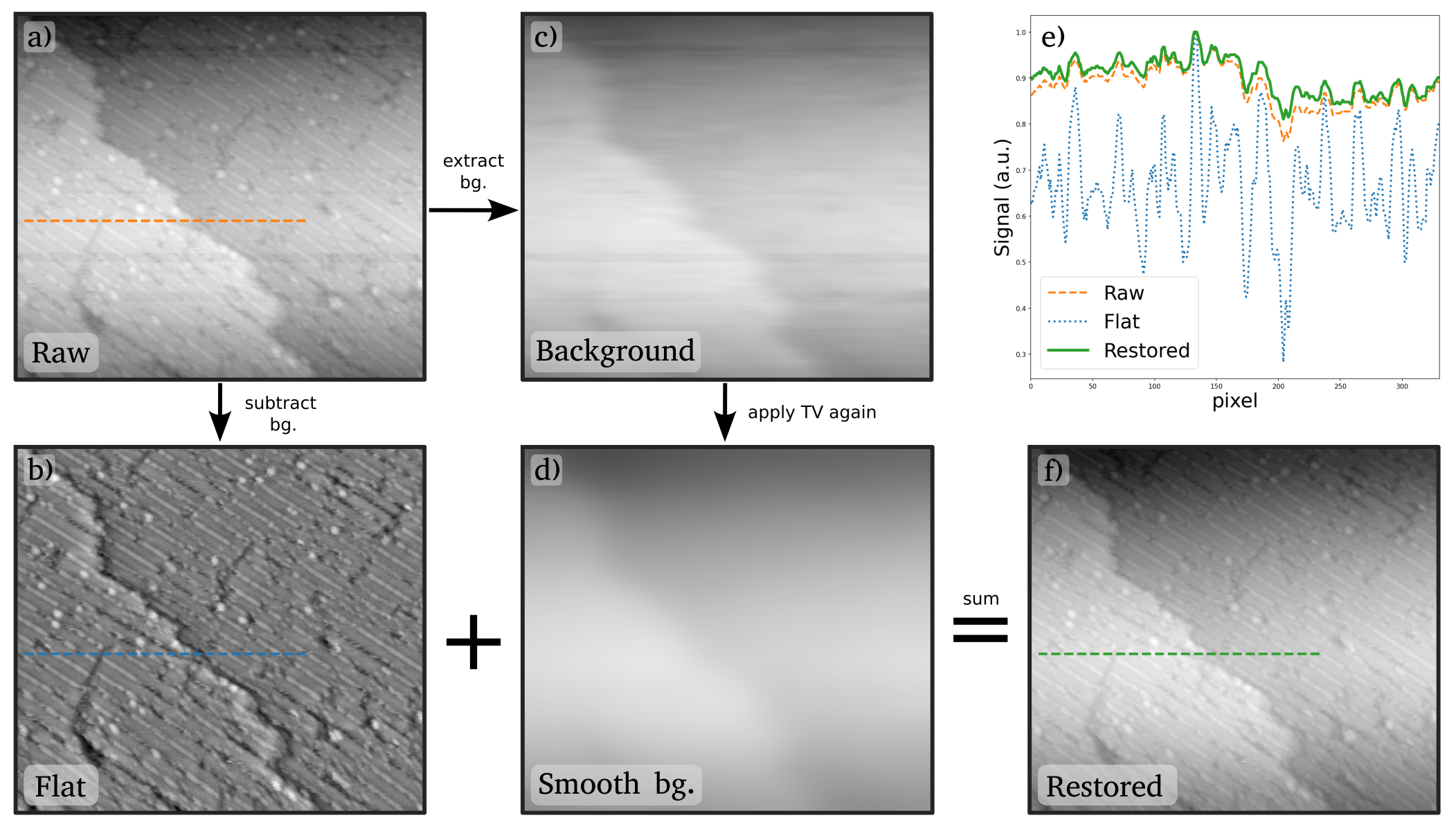}
    \caption{Restoration of the STM image of a (2×5)-reconstructed SrTiO$_3$(110) surface displayed in Fig.~\ref{fig:exp_images}\textcolor{red}{c}, following the steps described in the \textit{workflow} subsection, using the Huber-ROF method with $\lambda=0.05$ for both background extraction and background blurring. These steps are required when the image contains different terminations, terraces, or in general topographically equivalent areas with different contrasts. (a) raw image; (b) flat image obtained with background subtraction; (c) background extracted with TV minimization; (d) smooth background, obtained by applying TV denoising on the background shown in (c); (e) line profiles relative to raw, flat and denoised images; each line profile is normalized to the maximum signal intensity of the image they refer to, for comparison purposes; (f) restored image.
    }
    \label{fig:terraces}
\end{figure*}

\begin{figure*}[t]
    \centering
    \includegraphics[width=1.9\columnwidth]{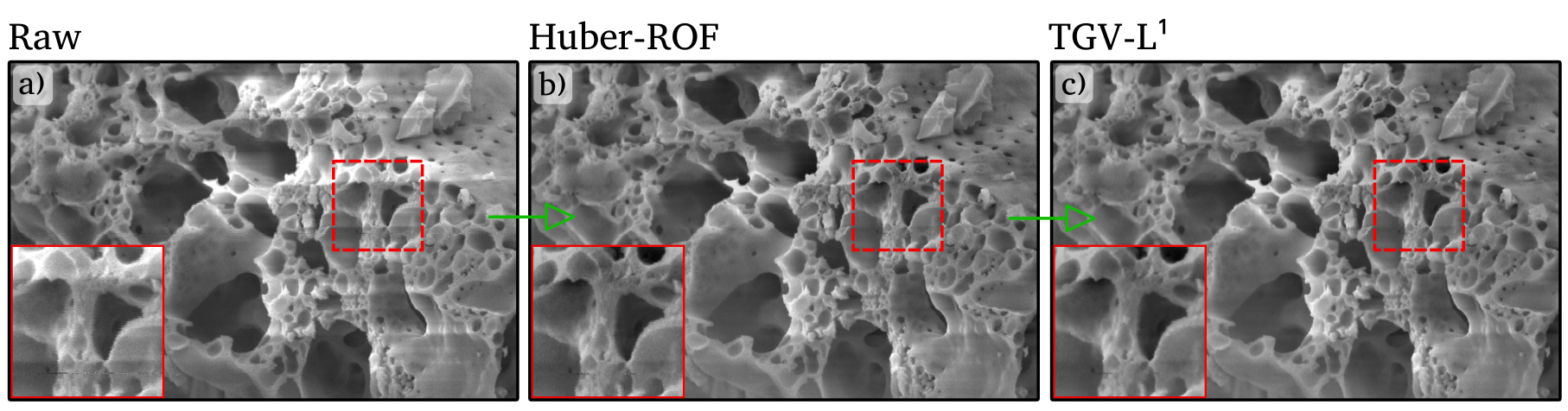}
    \caption{Restoration steps for the SEM image of MgAl$_2$O$_4$ shown in Fig.~\ref{fig:exp_images}\textcolor{red}{d}, with the steps described in the \textit{Workflow} subsection. The image is corrupted by bright, horizontal lines and low shot noise, which are removed with two different denoising steps. (a) Raw image and magnified region as inset; (b) image cleaned with Huber-ROF  ($\lambda=0.001$, $\alpha=0.005$), which removed most of the horizontal lines, with a magnified region as inset; (c) after applying Huber-ROF, TGV-$L^1$ has been applied to remove the low shot noise affecting the image, using $\lambda=1.5$.
    }
    \label{fig:MgAl2O4}
\end{figure*}

When terraces or terminations with different levels of contrast are present, the restoration process needs one extra step, as explained in the \textit{Methods} section and Fig.~\ref{fig:flowchart}. Figure~\ref{fig:terraces} shows the restoration of an STM image of a (2×5)-reconstructed SrTiO$_3$(110) surface using Huber-ROF with $\lambda=0.05$ for both background extraction and background blurring. In this case, extracting the background with TV minimization removes the contrast between the terraces, resulting in the image labeled as 'Flat'~(\ref{fig:terraces}\textcolor{red}{b}), where features belonging to different terraces or terminations display roughly the same intensity. The method we propose to recover the original contrast between terraces or terminations consists, with a second step, of applying TV on the extracted background~(\ref{fig:terraces}\textcolor{red}{c}) and obtaining what we label as 'Smooth background': this extra step aims to remove the horizontal lines from the background, obtaining a smoother version of it that preserves the contrast between terraces, as can be observed in panel~\ref{fig:terraces}\textcolor{red}{d}.
What is shown in Fig.~\ref{fig:terraces}\textcolor{red}{c} and Fig.~\ref{fig:terraces}\textcolor{red}{d} is comparable to applying a highpass and then a lowpass filter, respectively.
The line profiles associated with the raw, flat, and restored~(\ref{fig:terraces}\textcolor{red}{f}) images are shown in panel~\ref{fig:terraces}\textcolor{red}{e}. The raw image exhibits a modest contrast between bright features and the background but displays a distinct contrast between the terraces. In the flattened image, the contrast between terraces is almost absent, although the bright and dark features become more prominent. Incorporating the smoother background slightly reduces the overall contrast but significantly restores the original contrast between terraces. The dark, horizontal lines have been removed. However, dark shades in the upper and lower parts of the image persist.\\
Figure~\ref{fig:MgAl2O4} presents the restoration outcomes for a SEM image of MgAl$_2$O$_4$. In contrast to the previous example, this image not only features bright lines parallel to the scanning direction but also exhibits low shot noise. While background subtraction effectively addresses the horizontal lines (low-frequency signal), an additional step is required to eliminate the other noise component, whether it be shot noise, Gaussian noise, salt\&pepper noise, or a similar type.
Typically, Gaussian and shot noise can be removed via low-pass filtering, salt\&pepper noise via median filtering, and low-frequency background with high-pass filtering, but TV can cope with all of them depending on the setup of $\lambda$; Fig.~\textcolor{red}{S2} in the \textit{Supplementary material} shows a comparison between images cleaned with our proposed workflow, and with well-known filtering techniques implemented in ImageJ. The two methodologies, albeit based on different theoretical backgrounds, show reciprocal consistency.
As previously stated in Fig.~\ref{fig:lineplotLSMO}, Huber-ROF applied to the horizontal direction is optimal to remove the bright lines, and the results of background subtraction are shown in Figure~\ref{fig:MgAl2O4}\textcolor{red}{b}. The median filter applied to the vertical direction smoothed the features' edges by decreasing the small scan instability. Except for some bright areas, the lines have been removed. To eliminate noise present in the image, TGV-$L^1$ is subsequently applied to the image in panel \textit{b}), obtaining what is shown in panel \textit{c}): most of the leftover noise has been removed, and the edges have been preserved. We clarify that in such scenarion low shot noise is not strictly necessary to remove, but we decided to perform this additional step for demostration purposes of the well-known denoising capabilities of TV. The overall results show how the depth and details of the image are slightly reduced with respect to the raw image, and some bright areas are still present; the latter are flat, bright signals that do not encode physical information about the system, and their hypothetical removal will not reveal any hidden detail.

When images are heavily afflicted by shot noise or similar, it is sufficient to apply TV denoising to directly obtain the clean image \textit{u}.
As TV denoising is not the main focus of this work, but rather its image decomposition capabilities, we decided to show this scenario in the \textit{Supplementary material}, where we tested the denoising capabilities of TGV-L$^1$ against high shot noise present in a STEM image of graphene (Fig.~\textcolor{red}{S2b}) and simulated MoS$_2$ STEM images (more informations on Fig.~\textcolor{red}{S3}).


%% file: sections/conclusions.tex
This work presents a total variation-based workflow to restore microscopy images by performing background subtraction and denoising. Five different study cases have been proposed taken from AFM, STM, SEM, and STEM images. The Huber-ROF, TV-$L^1$, and TGV-$L^1$ methods have been applied.
For background subtraction, the Huber-ROF method is the most flexible one among the three since it allows with the sole control of the $\lambda$ parameter to filter out signals of different sizes, extracting the background from the raw image with great control. The TGV-$L^1$ method is confirmed to be the most edge-preserving one among the three, and it is suited for denoising but less for background extraction. TV-$L^1$ proved to be not enough sensitive to small features like atoms, making it unsuitable for background subtraction but still usable for texture smoothing (even though TGV-$L^1$ outperforms it) . \\
Our workflow mainly focuses on background subtraction, proven to be  particularly suitable for images displaying tip changes or sudden contrast variations due to tip-sample interactions, imaged as bright or dark lines parallel to the scanning direction. In the case of Gaussian noise or similar, obtaining a clean image directly from TV minimization is recommended, as shown in the case of the SEM image of MgAl$_2$O$_4$ and the simulated STEM images of MoS$_2$ in the supplementary materials. \\
A second processing step is required in cases when the image contains terraces or different terminations, which display different signal intensities: this is necessary to remove the small-scale noise from the background, ideally keeping the original contrast between terraces or terminations. The method proved effective but we have  not found a way to remove possible signal modulations not associated with the terraces or terminations.
In the third case we have discussed the image contains both horizontal bright or dark lines and Gaussian-like noise. The two-steps process consists first in removing the horizontal lines through background subtraction, then denoising the resulting image using preferably TGV-$L^1$. \\
The positive outcomes of this study suggest a wider degree of applicability of this workflow to other types of images. The code, alongside the results shown in this work, is publicly available on GitHub as a part of the AiSurf package~\parencite{corrias2023automated}, an open-access tool to clean and analyze microscopy images without requiring extensive knowledge of the user in the field. This workflow is conceived for single or multiple image denoising: it can be integrated into automated experimental workflows or used as a standalone code to denoise noisy image datasets.
For denoising purposes, many different methods have been proposed to automate the choice of $\lambda$~\parencite{goyal2020denoising, pan2019regularization, langer2017automated, selesnick2014tv}, however, to the best of our knowledge, there seem to be no suitable methods for background extraction; this might be due to the difficulty of giving a formal definition of background, which is mostly vague, case-dependent and not describable by means of a model. Developing a robust method to select $\lambda$ will be considered as a future outlook. The comparison with manually filtered images also shows the consistency of TV restoration with other techniques, which have shown equally-positive results.
\\